\documentclass[aps,floatfix,pra,preprint,showpacs]{revtex4-1}
\input{Qcircuit}
\usepackage{appendix,amsmath,amssymb,bbm,graphicx,nccmath,subfigure}

\newcommand{\norm}[1]{\left\lVert#1\right\rVert}
\newcommand{\Lim}[1]{\raisebox{0.5ex}{\scalebox{0.8}{$\displaystyle \lim_{#1}\;$}}}
\newtheorem{theorem}{Theorem} 

\begin{document}

\title{Efficient quantum circuits for continuous-time quantum walks on composite graphs}
\author{T. Loke and J. B. Wang}
\affiliation{School of Physics, The University of Western Australia, 6009 Perth, Australia}

\begin{abstract}

In this paper, we investigate the simulation of continuous-time quantum walks on specific classes of graphs, for which it is possible to fast-forward the time-evolution operator to achieve constant-time simulation complexity and to perform the simulation exactly, while maintaining $\mbox{poly}(\mbox{log}(n))$ efficiency. In particular, we discuss two classes of composite graphs, commuting graphs and Cartesian product of graphs, that contain classes of graphs which can be simulated in this fashion. This allows us to identify new families of graphs that we can efficiently simulate in a quantum circuit framework, providing practical and explicit means to explore quantum-walk based algorithms in laboratories.

\end{abstract}


\maketitle

\section{Introduction}

Quantum walks are currently a subject of intense theoretical and experimental investigation due to its established role in quantum computation and quantum simulation \cite{farhi_quantum_1998,kempe_quantum_2003,childs_universal_2009,childs_universal_2013,manouchehri_physical_2014}.  In fact, any dynamical simulation of a Hamiltonian system in quantum physics and quantum chemistry can be discretized and mapped onto a continuous-time quantum walk on specific graphs \cite{lloyd_universal_1996,aharonov_adiabatic_2003,berry_black-box_2012}. The primary difficulty of such a numerical simulation lies in the exponential scaling of the Hilbert space as a function of the system size, making real-world size problems intractable on classical computers. However, in order to run quantum walk-based algorithms on a quantum computer, we require an efficient quantum circuit that implements the required quantum walk. Some examples of quantum walk-based algorithms are searching for a marked element on a graph \cite{shenvi_quantum_2003,childs_spatial_2004,magniez_search_2011}, determining the relative importance of nodes in a graph \cite{paparo_google_2012,paparo_quantum_2013}, and testing graph isomorphism \cite{douglas_classical_2008}.

There are two distinct types of quantum walks: the continuous-time quantum walk (CTQW) \cite{farhi_quantum_1998} and the discrete-time quantum walk (DTQW) \cite{aharonov_quantum_1993}.  In previous studies, DTQWs on several classes of graphs have been shown to be efficiently implementable in terms of a sequence of quantum logic gate operations \cite{douglas_efficient_2009, jordan_efficient_2009,chiang_efficient_2010, loke_efficient_2012}. The implementation of the DTQW time-evolution operator $U(t) = (S \cdot C)^t$ for any $t \in \mathbb{R}_+$ is simplified by the fact that the time $t$ is discrete and the evolution is repetitive. Here $S$ and $C$ are the shift and coin operator, respectively.  As a result, if we can implement a single time-step $U = S \cdot C$ in a quantum circuit, the implementation for $U(t)$ can be generated by repeating the same circuit $t$ times.  Another property of DTQWs that is exploited in quantum circuit design is that the single time-step operator $U$ acts locally on the vertex-coin states encoding the graph. In other words, applying $U$ to the vertex-coin states associated with a particular vertex will only propagate the corresponding amplitudes to adjacent vertices, so vertices that are a distance of two or more apart do not affect each other in the single time-step. This means that the local structure of the graph is the primary consideration in implementing DTQWs.  Taking these into account, quantum circuits to implement DTQWs on sparse graphs \cite{jordan_efficient_2009,chiang_efficient_2010}, graphs with a high degree of symmetry, such as vertex-transitive graphs \cite{douglas_efficient_2009}, and certain types of non-degree-regular graphs \cite{loke_efficient_2012} have been constructed.

However for CTQWs, the above two properties of DTQWs do not carry over, which makes the design of quantum circuits substantially more difficult. First, the time-evolution operator $U(t)$ for CTQW is defined as a continuous function of $t \in \mathbb{R}_+$, requiring a time-dependent quantum circuit implementation. Second, the CTQW does not act locally on vertices - any two even distantly connected vertices on a graph will propagate amplitudes to each other in a CTQW. Consequently, the global structure of a graph must be taken into account in designing quantum circuits to implement CTQWs.

A substantial amount of work has been done on implementing CTQWs efficiently on quantum computers, typically considered under the more general problem of Hamiltonian simulation. Two classes of Hamiltonians are considered separately: sparse Hamiltonians \cite{childs_exponential_2003,berry_efficient_2007,kothari_efficient_2010,childs_simulating_2011,wiebe_simulating_2011,berry_hamiltonian_2015} and dense Hamiltonians \cite{childs_limitations_2009, kothari_efficient_2010}. Simulation of an $n$-dimensional Hamiltonian $H$ is said to be efficient if there exists a quantum circuit using at most $\mbox{poly}(\mbox{log}(n),\norm{Ht},1/\epsilon)$ one- and two- qubit gates (where $\mbox{poly}(\ldots)$ denotes a polynomial scaling in terms of the listed parameters) that approximates the evolution operator $\mbox{exp}(-i t H)$ with error at most $\epsilon$ \cite{kothari_efficient_2010}.  Here $\norm{..}$ denotes the spectral norm of a matrix. 

Under this definition of efficiency, the complexity of quantum circuits would, in general, scale at least linearly with the time $t$. Given the periodic nature of a unitary system, this is undesirable although necessary in general, due to the no-fast-forwarding theorem \cite{berry_efficient_2007}:

\begin{theorem}
	\textnormal{(No-fast-forwarding theorem)} For any positive integer $N$ there exists a row-computable sparse Hamiltonian with $\norm{H}=1$ such that simulating the evolution of $H$ for time $t=\pi N/2$ within precision $1/4$ requires at least $N/4$ queries to $H$.
\end{theorem}

As such, simulation of general sparse Hamiltonians in sublinear time is not possible (this result has been extended to non-sparse Hamiltonians as well in \cite{childs_limitations_2009}). However, there still exists classes of Hamiltonians that can be simulated in sublinear (or even constant) time, i.e. Hamiltonians for which their time-evolution can be fast-forwarded, as pointed out in \cite{childs_limitations_2009}.

In this paper, we identify classes of graphs for which their Hamiltonian (given by the adjacency matrix) can be simulated efficiently in constant time (i.e. the complexity of the quantum circuit does not scale with the parameter $t$). We also focus on exact simulations of these graphs, that is, $\epsilon=0$. Hence, in the rest of this paper, we term a quantum circuit implementation of a CTQW as efficient if it uses at most $\mbox{poly}(\mbox{log}(n))$ one- and two- qubit gates to implement the CTQW time-evolution operator $U(t)=\mbox{exp}(-i t H)$ of a graph on $n$ vertices with $\epsilon=0$, with the complexity being $t$-independent. This paper mirrors the efforts of previous studies in the DTQW literature to implement exactly the DTQW step operator for certain classes of graphs \cite{douglas_efficient_2009,loke_efficient_2012}, where it was found that only graphs that are highly symmetric are amenable to exact efficient implementations. This study is also motivated by the need for highly efficient quantum circuit implementations of CTQWs, since additional operations or a complete redesign of quantum circuits for different values of $t$ would be prohibitively expensive given current experimental capabilities.

There are a select few classes of graphs for which their CTQW can be fast-forwarded to obtain an efficient quantum circuit implementation. In particular, it has been pointed out previously that the glued tree \cite{childs_exponential_2003}, complete graph, complete bipartite graph and star graph \cite{childs_relationship_2010} can be simulated efficiently using diagonalization. Recently, the class of circulant graphs (with some restrictions) have also been identified as being efficiently implementable \cite{qiang_efficient_2015}, since circulant graphs can be diagonalized using the quantum Fourier transform.



The paper is organized as follows. In section \ref{sec:diag}, we review the background theory of CTQWs on graphs, and discuss diagonalization of matrices, which maps to quantum circuit implementations that have time-independent complexity. We discuss commuting graphs in section \ref{sec:commg}, which yield new graphs that can be efficiently implemented using known implementations of the subgraphs. In section \ref{sec:cartp}, we discuss the Cartesian product of graphs, which further extends the classes of graphs we can efficiently implement. We then draw our conclusions in section \ref{sec:conc}.

\section{Background theory}
\label{sec:diag}

Consider a general undirected graph $G(V,E)$, with vertex set $V=\left\{v_1,v_2,v_3,\ldots\right\}$ and edge set $E=\left\{(v_i,v_j),(v_k,v_l),\ldots\right\}$ being unordered pairs connecting the vertices. Suppose that $G$ has $n$ vertices. The $n$-by-$n$ adjacency matrix $A$ is defined as $A_{jk} = 1$, if $(v_j,v_k) \in E$ and $0$ otherwise. For an undirected graph, its adjacency matrix $A$ is symmetric. The degree $d_i$ of a vertex $v_i$ in an undirected graph is the number of undirected edges connected to $v_i$, which we denote by $\mbox{deg}(A)_{v_i} = d_i$. A degree-regular graph with degree $d$ is then a graph that has $\mbox{deg}(A)_{v_i} = d \mbox{ } \forall i = 1,\ldots,N$, i.e. every vertex has the same degree $d$.

The CTQW is described by a state vector $\ket{\psi(t)}$ in the Hilbert space of dimension $n$ spanned by the orthonormal basis states $\{ \ket{1}, \ket{2}, \ldots, \ket{n} \}$ corresponding to vertices in the graph.  The time-evolution of the state $\ket{\psi(t)}$ is governed by the time-dependent Schr{\"o}dinger equation
\begin{equation}
i \frac{d}{dt} \ket{\psi(t)} = H \ket{\psi(t)},
\end{equation}
where the Hamiltonian $H$ is a Hermitian operator, i.e. $H=H^\dag$. The formal solution to this equation is $\ket{\psi(t)} = U(t) \ket{\psi(0)}$, where $U(t) = \mbox{exp}(-i t H)$ is the time-evolution operator. Choice of $H$ varies in the literature between $H=\gamma \left( D - A \right)$ \cite{childs_example_2002,kempe_quantum_2003} and $H = \gamma A$ \cite{farhi_quantum_1998, chakrabarti_design_2012}, where $\gamma$ is the hopping rate per edge per unit time and $D$ is an $n$-by-$n$ (diagonal) degree matrix  defined by $D_{ij} = d_i \delta_{ij}$, where $\delta_{ij}$ is the Kronecker delta. For degree-regular graphs, the only difference between the two choices is a global phase factor and a sign flip in $t$, which does not change observable quantities \cite{ahmadi_mixing_2003}. However, the two choices will result in different dynamics for non-degree-regular graphs \cite{childs_spatial_2004}. In this paper, we use $H = \gamma A$.

It is well-known that for a Hermitian matrix $H$, the spectral theorem guarantees that $H$ can be diagonalized using its eigenbasis, that is $H = Q^\dag \Lambda Q$~\cite{hogben_handbook_2006}. 
Here $Q$ is a unitary matrix whose column vectors are eigenvectors of $H$, and $\Lambda$ is a diagonal matrix of eigenvalues of $H$, which are all real and whose order is determined by the order of the eigenvectors in $Q$. From this, we can express the time-evolution operator as 
$U(t) = Q^\dag \mbox{exp}(-i t \Lambda) Q$.

The diagonalization approach confines the time-dependence of $U(t)$ to the diagonal matrix $\mbox{exp}(-i t \Lambda)$, which can be readily implement by a sequence of at most $n$ controlled-phase gates with phase values being linear functions of $t$. Experimentally, this corresponds to a sequence of tunable controlled-phase gates, where the phase values are determined by $t$. 

Two quantum gates that we will use heavily in the following sections are the Hadamard gate:

\begin{equation}
\mathcal H = \frac{1}{\sqrt{2}}
\begin{pmatrix} 
1&1 \\ 1&-1 
\end{pmatrix},
\end{equation}

and the general 2-phase rotation gate:

\begin{equation}
R(\theta_1,\theta_2) = 
\begin{pmatrix} 
e^{i \theta_1}&0 \\ 
0&e^{i \theta_2} 
\end{pmatrix}.
\end{equation}


In principle, a quantum circuit implementation for $U(t)$ that has time-independent complexity can always be obtained by using a general quantum compiler (as discussed in \cite{chen_qcompiler:_2013,loke_optqc:_2014}) to obtain a quantum circuit implementation for $Q$ and $Q^\dag$ - however this is almost never efficient, since such methods typically scale exponentially in terms of complexity. In order to be able to implement $U(t)$ efficiently as a whole, we require that at most $\mbox{poly}(\mbox{log}(n))$ one- and two- qubit gates are used in implementing $Q$, $\mbox{exp}(-i t \Lambda)$ and $Q^\dag$ individually (which is not always possible, as per the no-fast-forwarding theorem). In the following sections, we will cover some classes of graphs that satisfy this criteria.

\section{Commuting graphs}
\label{sec:commg}

Suppose we have two matrices $H_1$ and $H_2$. In general, when $H_1$ and $H_2$ do not commute, their sum can be simulated by the Lie product formula \cite{childs_quantum_2004}
\begin{equation}
\mbox{exp}(-i t (H_1+H_2)) = \Lim{m \rightarrow \infty} \left( \mbox{exp}(-i t H_1 / m) \mbox{ exp}(-i t H_2 / m) \right)^m
\end{equation}
which, in practice, requires high-order approximations to achieve a bounded error that scales with $t$. However, in the case where $H_1$ and $H_2$ commute, we can write the expression exactly as
\begin{equation}
\mbox{exp}(-i t (H_1+H_2)) = \mbox{exp}(-i t H_1) \mbox{ exp}(-i t H_2).
\label{eqn:ucomm}
\end{equation}
Taking $H_1 = \gamma A$ and $H_2 = \gamma B$, where $\gamma$ is constant, $A$ and $B$ are the adjacency matrices of two commuting graphs, i.e. $\left[ A , B \right] = 0$. It follows that if the individual time-evolution operators $\mbox{exp}(-i t \gamma A)$ and $\mbox{exp}(-i t \gamma B)$ can be efficiently implemented, then the time-evolution operator for the graph $A+B$, that is, $\mbox{exp}(-i t \gamma (A+B))$, can also be efficiently implemented, provided $\left[ A , B \right] = 0$.

The general criteria for commuting graphs is studied in \cite{nehaniv_construction_2014}. One particular class of graphs is the interdependent networks, defined by $$ A = \begin{pmatrix} A_1&0 \\ 0 & A_2 \end{pmatrix}  \textrm{ and ~}  B = \begin{pmatrix} 0 & B_0 \\ B_0^T&0 \end{pmatrix}, $$
where the interlink graph $B$ connects two subgraphs $A_1$ and $A_2$, which are both symmetric, i.e. $A_1 = A_1^T$ and $A_2 = A_2^T$. In this instance, the condition for commutativity becomes
\begin{equation}
A_1 B_0 = B_0 A_2.
\label{eqn:commc}
\end{equation}
Suppose $Q_1$ and $Q_2$ diagonalize $A_1$ and $A_2$ respectively, we have 
$$
\Lambda_1 = Q_1^\dag A_1 Q_1  \textrm{~ and ~}  \Lambda_2 = Q_2^\dag A_2 Q_2 . 
$$
Then the following matrix 
$$ Q = \begin{pmatrix} Q_1 & 0 \\ 0 & Q_2 \end{pmatrix} $$ 
diagonalizes $A$, and give the eigenvalue matrix 
$$ \Lambda = \begin{pmatrix} \Lambda_1 & 0 \\ 0 & \Lambda_2 \end{pmatrix} .$$ 
Suppose $B_0$ is diagonalized by $Q_0$, i.e. $ \zeta_0 = Q_0 B_0 Q_0^\dag $, then it can be shown that if $B_0 = B_0^T$ (namely a symmetric interconnection), the diagonalizing matrix for $B$ is 
$$Q' = \frac{1}{\sqrt{2}} \begin{pmatrix} Q_0 & Q_0 \\ Q_0 & -Q_0 \end{pmatrix} = \mathcal H \otimes Q_0, $$ 
where $\mathcal H $ is the Hadamard matrix as defined above.  The corresponding eigenvalue matrix 
$$ \zeta = \begin{pmatrix} \zeta_0 & 0 \\ 0 & -\zeta_0 \end{pmatrix} = \sigma_z \otimes \zeta_0, $$
where $\sigma_z$ is the Pauli-z matrix. Hence, in the case where $\left[ A , B \right] = 0$, we expand the CTQW time-evolution operator as
\begin{equation}
\mbox{exp}(-i t (H_1+H_2)) = Q^\dag \mbox{exp}(-i t \Lambda) Q \mbox{ } Q'^\dag \mbox{exp}(-i t \zeta) Q' .
\end{equation}

\begin{figure}[htp]
	\centering
	\subfigure[]{\includegraphics[scale=0.2]{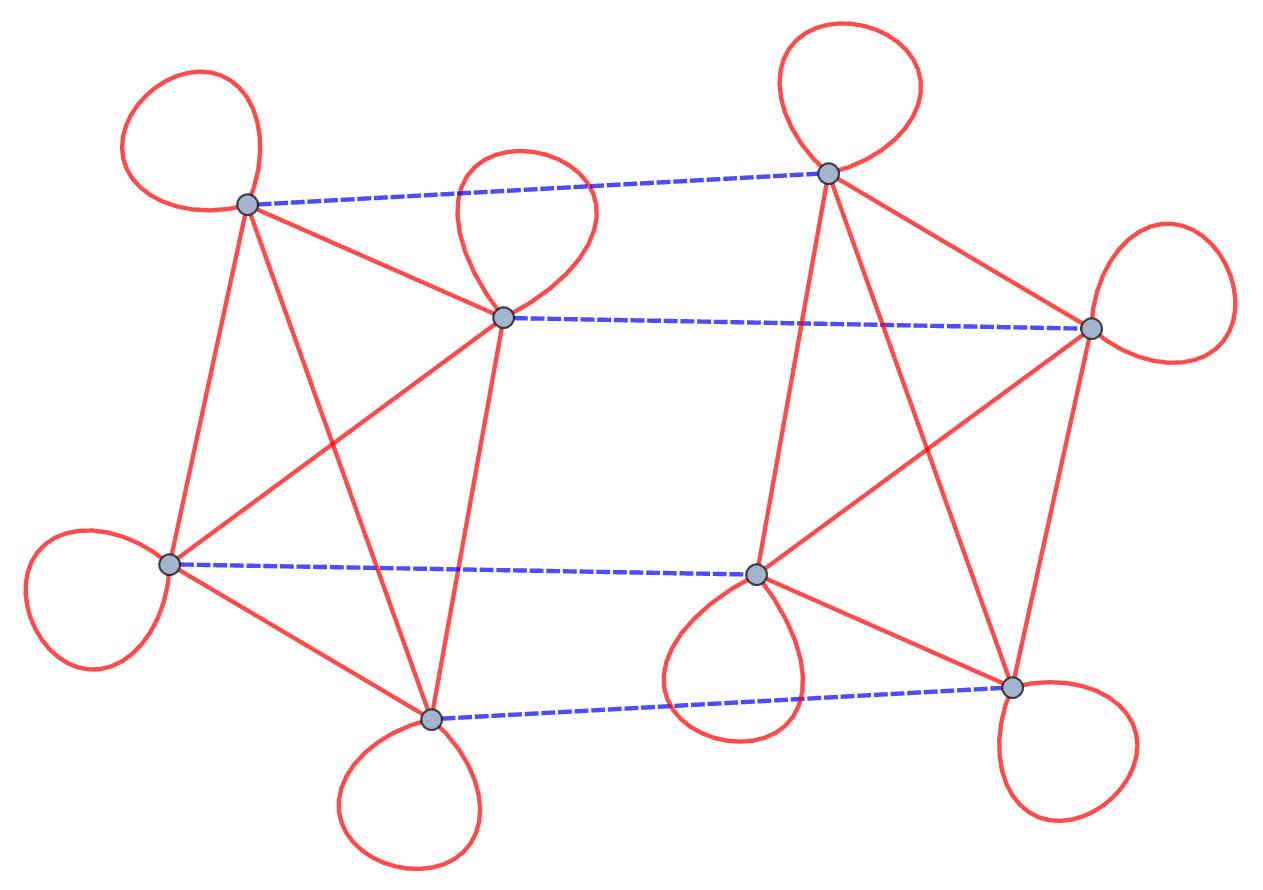}}
	\subfigure[]{\includegraphics[scale=0.32]{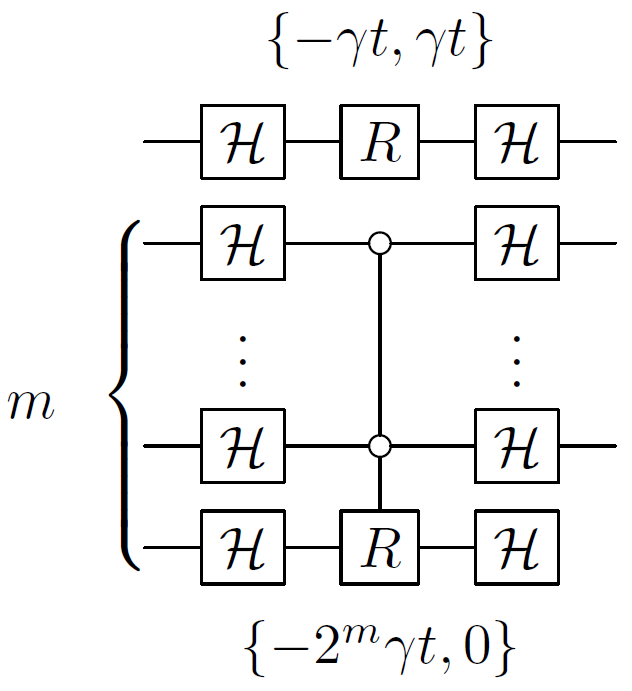}}
	\caption{(a) Two disjoint $K_{2^m}$ graphs with identity interconnections (solid red lines and dashed blue lines indicate edges belonging to $A$ and $B$ respectively), where $m=2$; (b) the corresponding quantum circuit implementation for the CTQW time-evolution operator on this graph.}
	\label{fig:K4IIC}
\end{figure}

Next, we examine some explicit examples of interdependent networks in which $\left[ A , B \right] = 0$ is satisfied and Eq.~(\ref{eqn:commc}) holds. One special case is that of identity interconnections between two disjoint copies of a graph with $n$ vertices, namely $A_1 = A_2$ and $B_0=I_n$. The diagonalizing matrices for $A$ and $B$ are, respectively
$$ Q = \begin{pmatrix} Q_1 & 0 \\ 0 & Q_1 \end{pmatrix} = I_2 \otimes Q_1 \textrm{~ and ~} Q' = \mathcal H \otimes I_n, $$
giving the eigenvalue matrices 
$$\Lambda = I_2 \otimes \Lambda_1 \textrm{~ and ~} \zeta = \sigma_z \otimes I_n .$$
Hence, if we are able to implement $A_1$ efficiently, it follows that the interdependent network with $A_1 = A_2$ and $B_0=I_n$ can be implemented efficiently. An equivalent result can be achieved by noting that $A+B=\sigma_x \oplus A_1$, where $\sigma_x$ is the Pauli-x matrix and $\oplus$ denotes the Cartesian product (defined in section \ref{sec:cartp}), and then applying the methods of section \ref{sec:cartp}. Fig.~\ref{fig:K4IIC} shows one class of graphs (complete graphs with identity interconnections) that can be constructed as above, together with its corresponding circuit implementation.

\begin{figure}[htp]
	\centering
	\subfigure[]{\includegraphics[scale=0.3]{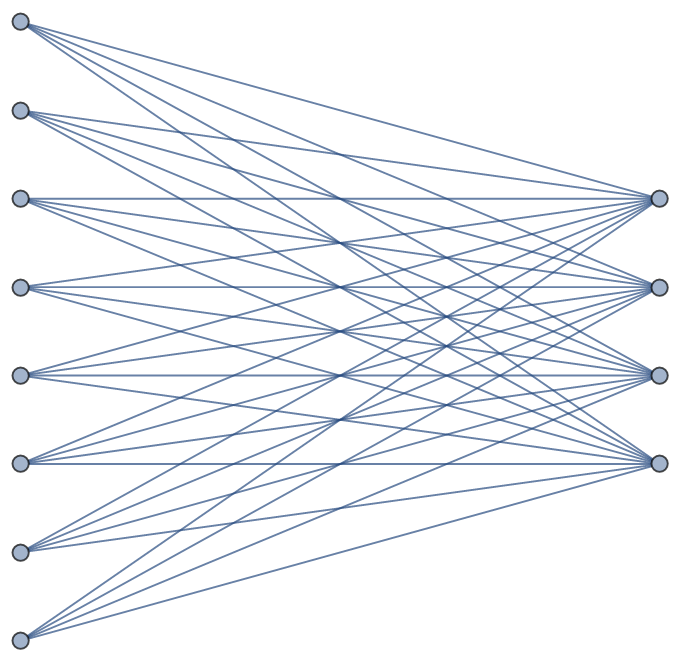}\qquad}
	\subfigure[]{\includegraphics[scale=0.22]{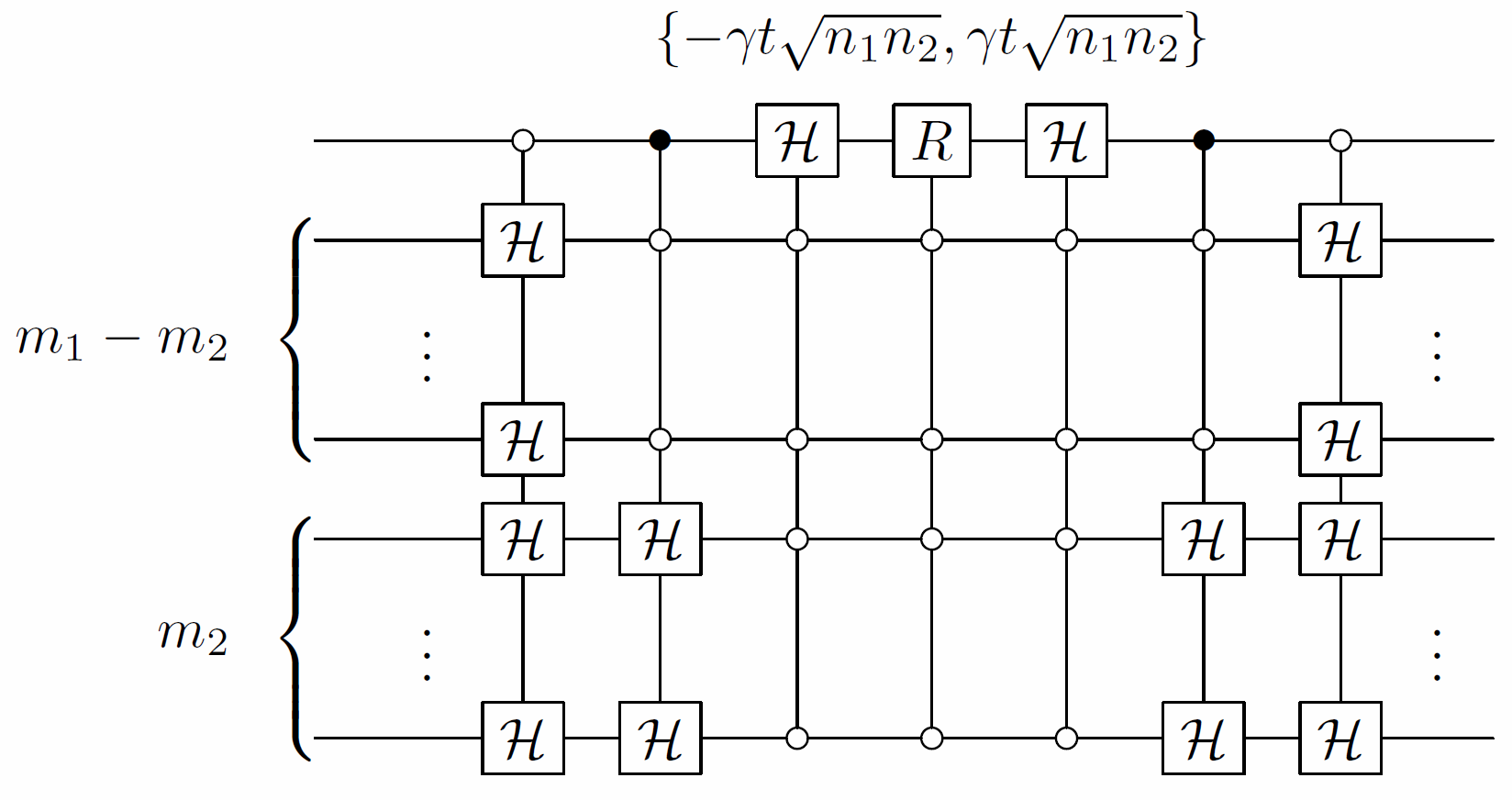}\label{fig:kn1n2cir}}
	\caption{(a) An example of the complete bipartite graph $K_{n_1,n_2}$, where $n_1=8$ and $n_2=4$; (b) quantum circuit implementation of the CTQW time-evolution operator for the complete bipartite graph $K_{n_1,n_2}$, where $n_1 = 2^{m_1}$ and $n_2 = 2^{m_2}$.}
	\label{fig:Kn1n2}
\end{figure}

Another case where $\left[ A , B \right] = 0$ is using complete interconnections between two disjoint degree-regular graphs (with $n_1$ and $n_2$ vertices respectively) of same degree.  That is, $\mbox{deg}\left(A_1\right)=\mbox{deg}\left(A_2\right)=d$ and $B_0=J_{n_1,n_2}$, where $J_{n_1,n_2}$ is the $n_1$-by-$n_2$ matrix with all 1's. Although in general $B_0 \neq B_0^T$, the interconnection matrix 

\begin{equation}
B = \begin{pmatrix} 0 & J_{n_1,n_2} \\ J_{n_2,n_1}&0 \end{pmatrix}
\label{eqn:interj}
\end{equation}  

can still be diagonalized easily in the case where $n_1=2^{m_1}$ and $n_2=2^{m_2}$, for non-negative integers $m_1$ and $m_2$. For convenience, we assume $n_1 \geq n_2$, and note that $B$ is the complete bipartite graph $K_{n_1,n_2}$. The diagonalization operator for $B$ can be written mathematically as
\begin{align}
Q' = & \left( I_{2^{m_1+1}} + \left( H - I_2 \right) \otimes P_0^{\otimes m_1} \right) \left( I_{2^{m_1+1}} + P_0 \otimes \left( H^{\otimes m_1} - I_{2^{m_1}} \right) + \right. \nonumber \\
& \left. P_1 \otimes P_0^{\otimes (m_1-m_2)} \otimes \left( H^{\otimes m_2} - I_{2^{m_2}} \right) \right),
\end{align}
where $P_0 = \ket{0}\bra{0}$ and $P_1 = \ket{1}\bra{1}$ are the 2-dimensional projection operators. The corresponding eigenvalue matrix of $B$ is then given by 
\begin{equation}
\zeta = \mbox{diag}\left(\left\{ (+\sqrt{n_1 n_2})^1, 0^{n_1-1},(-\sqrt{n_1 n_2})^1,0^{n_2-1} \right\}\right).
\end{equation}
Fig.~\ref{fig:Kn1n2} shows the complete bipartite graph $K_{n_1,n_2}$ together with its corresponding quantum circuit implementation. As a corollary, the star graph $S_{2^m+1}$ can also be implemented using the same method, since $S_{2^m+1}$ is equivalent to $K_{2^m,1}$.

\begin{figure}[htp]
	\centering
	\subfigure[]{\includegraphics[scale=0.22]{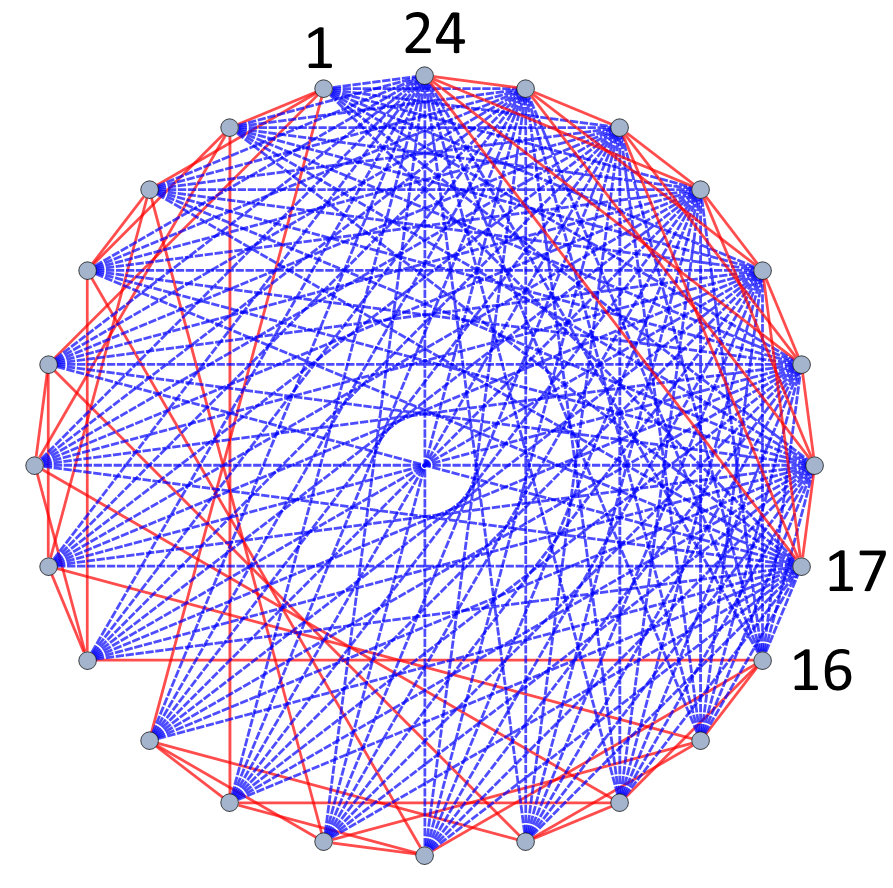}\qquad}
	\subfigure[]{\includegraphics[scale=0.25]{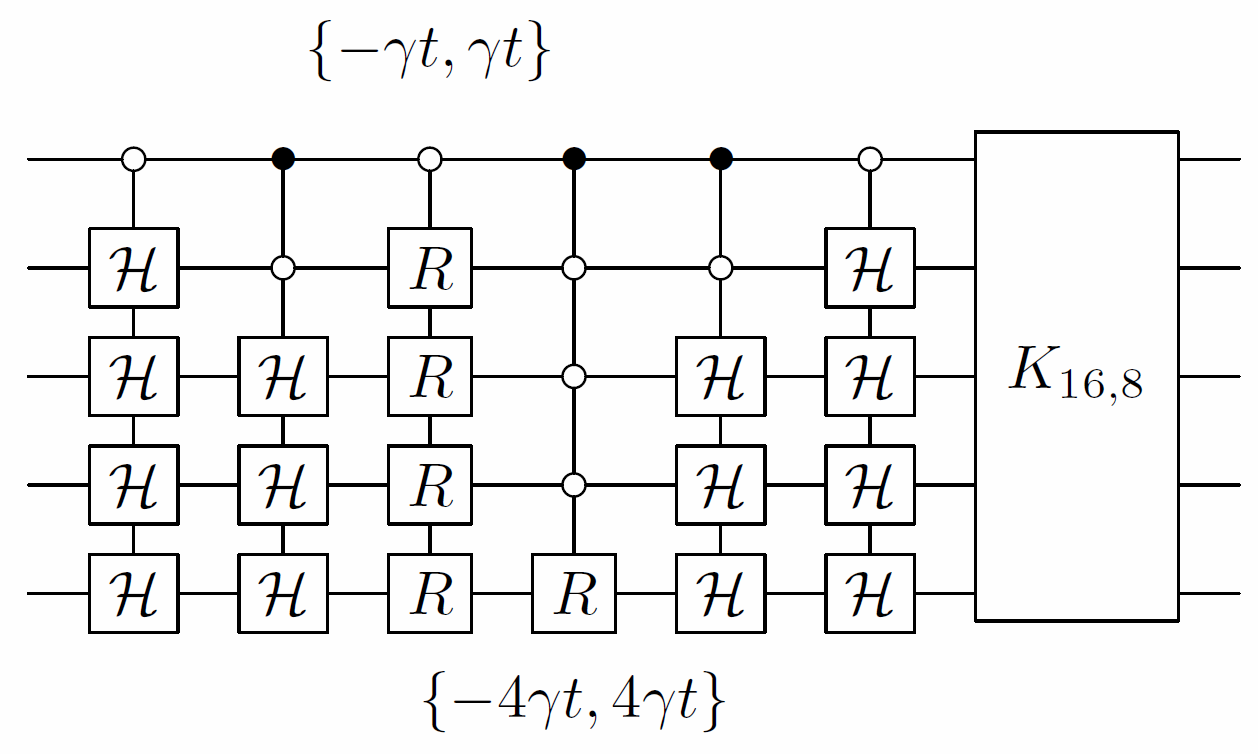}}
	\caption{(a) Disjoint $Q_4$ and $K_{4,4}$ with complete interconnections (solid red lines and dashed blue lines indicate edges belonging to $A$ and $B$ respectively); (b) its corresponding quantum circuit implementation. 
	}
	\label{fig:Q4Kh8}
\end{figure}

Hence, if we have degree-regular graphs $A_1$ and $A_2$ satisfying 

\begin{equation}
\mbox{deg}\left(A_1\right)_v=\mbox{deg}\left(A_2\right)_v=d \mbox{ } \forall v \in V,
\label{eqn:degreg}
\end{equation} 

where $n_1=2^{m_1}$ and $n_2=2^{m_2}$, which can both be efficiently implemented, then it follows that the interdependent network built from $A_1$, $A_2$ and $B_0=J_{n_1,n_2}$ can be efficiently implemented. Fig.~\ref{fig:Q4Kh8}(a) gives an example of this kind of graphs, where vertices 1-16 belong to the $Q_4$ graph (hypercube graph of dimension 4 - refer to section \ref{sec:cartp}), and 17-24 belong to the $K_{4,4}$ graph.  The quantum circuit implementation of the composite graph shown in Fig.~\ref{fig:Q4Kh8}(a) is given by Fig.~\ref{fig:Q4Kh8}(b), where the $K_{16,8}$ circuit is already described above and given by Fig.~\ref{fig:kn1n2cir}. Note that in general, the $K_{n_1,n_2}$ graph and by extension, the resulting interdependent network with complete interconnections is not degree-regular - so this provides an example of a class of graphs that is not degree-regular but still has an efficient quantum circuit implementation for the CTQW time-evolution operator.

%

\section{Cartesian product of graphs}
\label{sec:cartp}

Given two matrices $H_1$ and $H_2$ of dimension $n_1 \times n_1$ and $n_2 \times n_2$ respectively, the Cartesian product of $H_1$ and $H_2$ are given by
\begin{equation}
	H_1 \oplus H_2 = H_1 \otimes I_{n_2} + I_{n_1} \otimes H_2,
\end{equation}
which is a matrix of dimension $n_1n_2 \times n_1n_2$. In particular, if we define $H = H_1 \oplus H_2$, we have
\begin{equation}
	\mbox{exp}(-i t H) = \mbox{exp}(-i t H_1) \otimes \mbox{exp}(-i t H_2)
\end{equation}
or, more compactly, $U(t)=U_1(t) \otimes U_2(t)$.  Again we set $H_1 = \gamma A$ and $H_2 = \gamma B$, i.e. the Hamiltonians $H_1$ and $H_2$ correspond to graphs $A$ and $B$ respectively. This implies that if we have an efficient quantum circuit implementation for the individual CTQW time-evolution operators on the graphs $A$ and $B$, the implementation for $U(t)$ (which is the time-evolution operator that corresponds to the CTQW on the graph $A \oplus B$) is easily formed by stacking the individual quantum circuit implementations in parallel. As a corollary, in the case where $A = B$, then $H = A \oplus A$ corresponds to the Hamiltonian for the non-interacting two-particle quantum walk on $A$.

\begin{figure}[htp]
	\centering
	\subfigure[]{\includegraphics[scale=0.2]{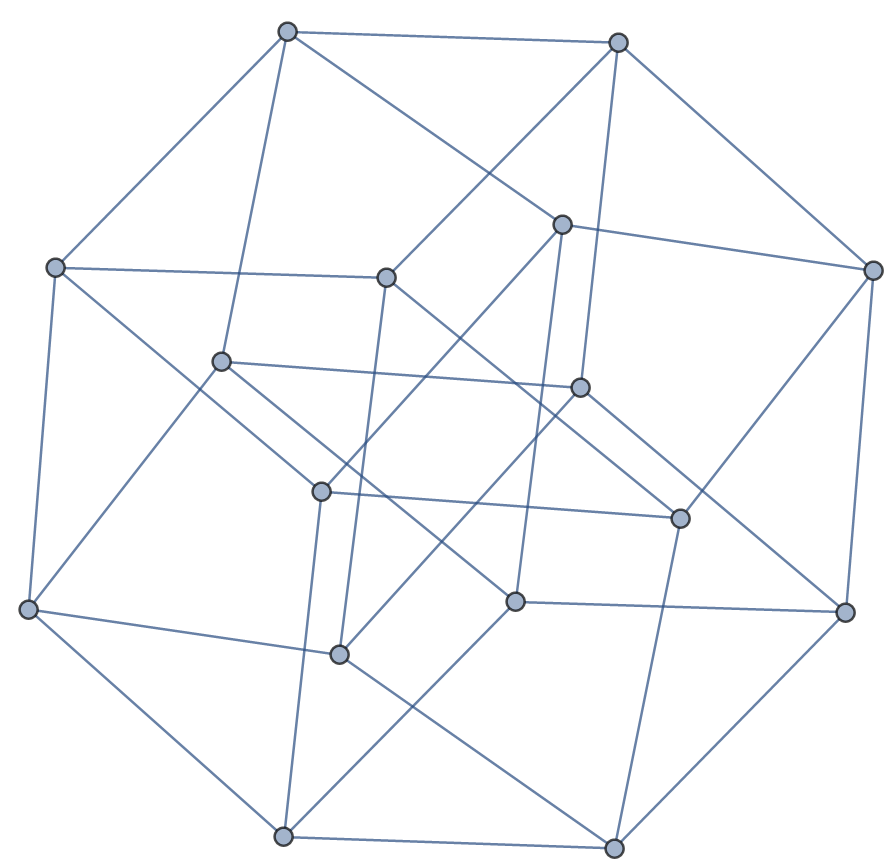}\qquad}
	\subfigure[]{\includegraphics[scale=0.25]{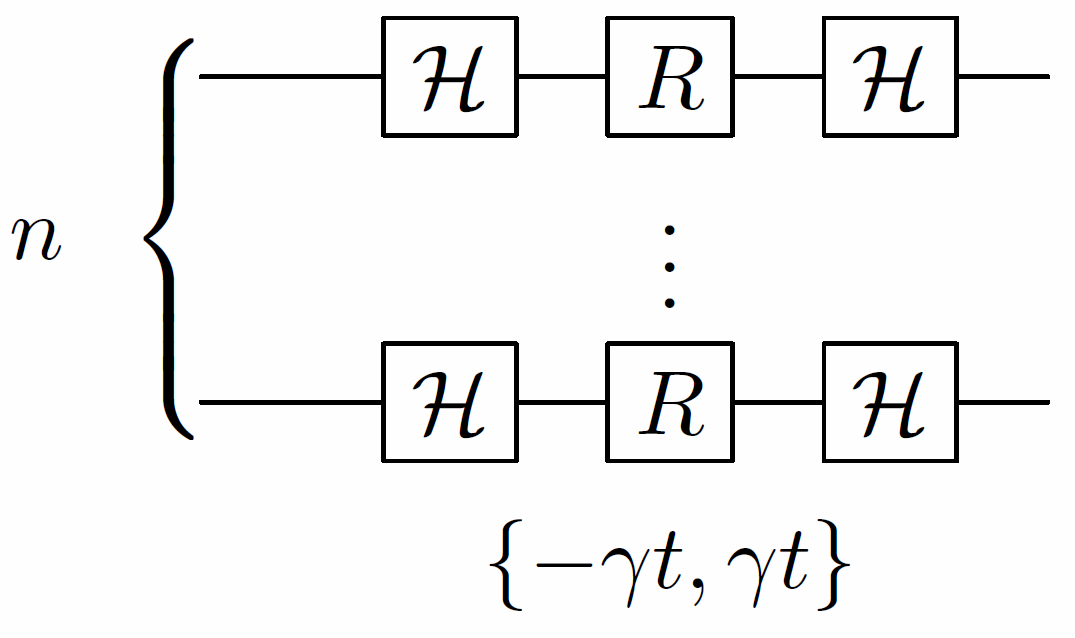}}
	\caption{(a) The hypercube graph $Q_n$, where $n=4$; (b) its corresponding quantum circuit implementation for the CTQW time-evolution operator on $Q_n$.}
	\label{fig:Qn}
\end{figure}

One particular class of graphs that is constructed using the Cartesian product of graphs is the hypercube $Q_n$. Given the path graph $P_2$ of length 2 with adjacency matrix $ \begin{pmatrix} 0&1 \\ 1&0 \end{pmatrix} $, $Q_n$ is constructed as $Q_n = P_2^{\oplus n}$, i.e. it is the Cartesian product of $n$ copies of $P_2$ \cite{harary_survey_1988}. As such, $Q_n$ is a graph with $2^n$ vertices and degree-regular with degree $n$. $P_2$ is diagonalizable using the Hadamard matrix $\mathcal{H}$, giving its eigenvalue matrix $ \Lambda_{P_2} = \begin{pmatrix} 1&0 \\ 0&-1 \end{pmatrix} $. Fig.~\ref{fig:Qn} shows the hypercube graph $Q_n$ with its corresponding quantum circuit implementation. 

\begin{figure}[htp]
	\centering
	\subfigure[]{\includegraphics[scale=0.25]{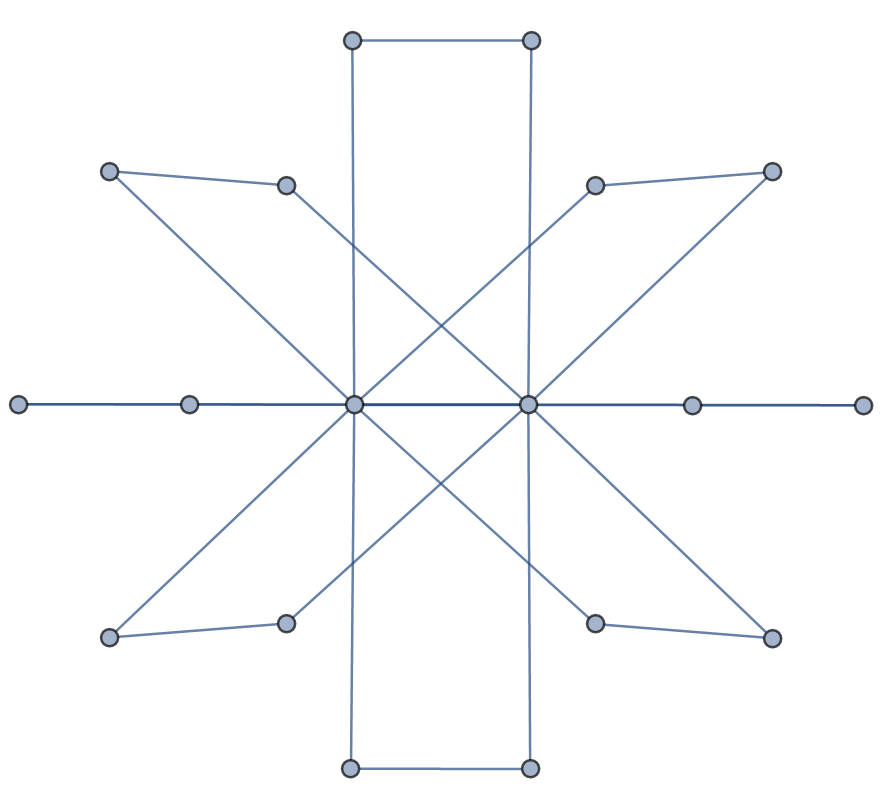}\qquad}
	\subfigure[]{\includegraphics[scale=0.25]{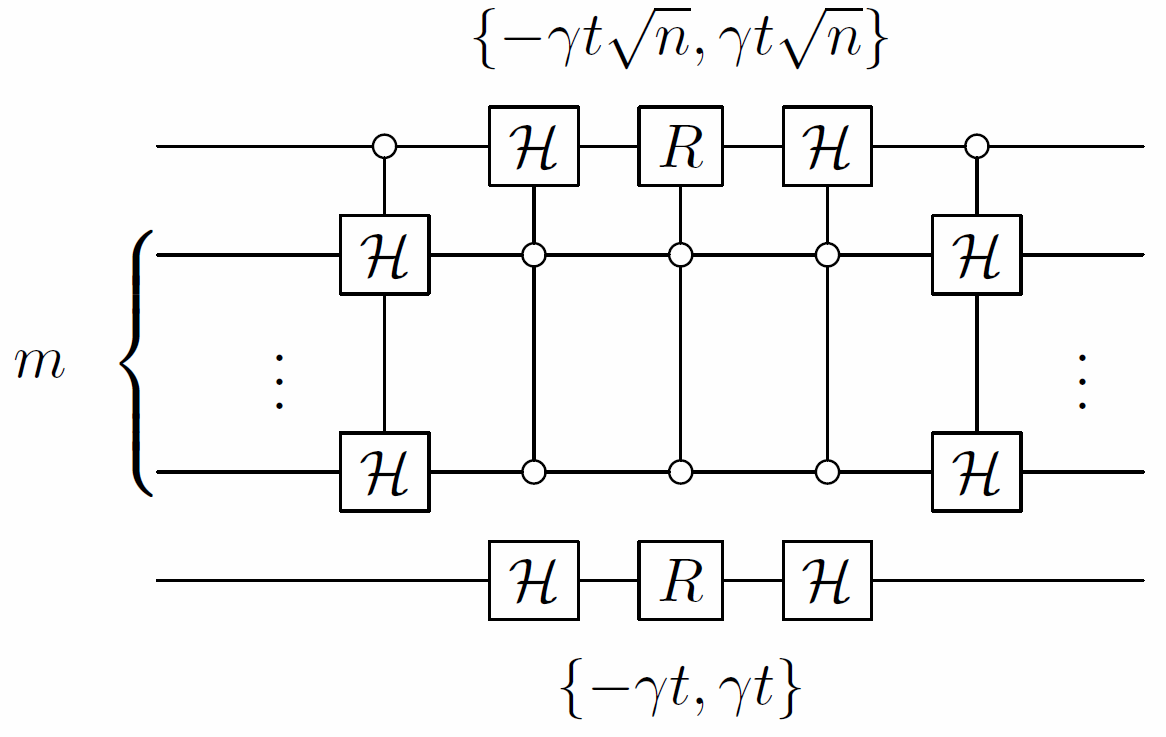}}
	\caption{(a) The book graph $B_n$, where $n=8$; (b) corresponding quantum circuit implementation for the CTQW time-evolution operator on the book graph for $n = 2^m$.}
	\label{fig:Bn}
\end{figure}

Another example of this class of graphs is the book graph $B_n$ \cite{gallian_dynamic_1997}, which is constructed as $B_n = S_{n+1} \oplus P_2 $, where $S_{n+1}$ is the star graph on $n+1$ vertices.  As we have discussed in section \ref{sec:commg}, $S_{n+1}$ can be implemented efficiently in a quantum circuit if $n=2^m$ for some non-negative integer $m$ - hence book graphs of the form $B_{2^m}$ can be efficiently implemented as well, as shown in Fig.~\ref{fig:Bn}.

\section{Conclusion}
\label{sec:conc}

In summary, we have shown that using diagonalization, we can fast-forward the simulation of some Hamiltonians corresponding to graphs in an efficient manner for some classes of composite graphs  (commuting graphs and Cartesian products of graphs) to achieve constant-time complexity with $\epsilon=0$. The quantum circuit implementations presented here are eminently useful in experimentally implementing CTQWs, since new classes of graphs can be simulated using existing graphs with minimal increase in complexity.




%

\end{document}